\documentclass [12pt]{article}
\def\maj#1{\ifmmode\mbox{\usefont{U}{msb}{m}{n}#1}\else{\usefont{U}{msb}{m}{n}#1}\fi}
\def\v#1{\mathbf{#1}}

\begin{document}

\title{\textbf{Novel approach to nonlinear susceptibility}}
\author{Monique COMBESCOT, Odile BETBEDER-MATIBET 
 \\ \small{\textit{GPS, Universit\'e Pierre et Marie
Curie and Universit\'e Denis Diderot,
CNRS,}}\\ \small{\textit{Campus Boucicaut, 140 rue de
Lourmel, 75015 Paris, France}}
\\ Kikuo CHO and Hiroshi AJIKI\\
\small{\textit{Graduate School of Engineering
Science, Osaka University,}}\\
\small{\textit{Machikaneyama, Toyonaka
560-8531, Japan}}}
\date{}
\maketitle

\begin{abstract}
The calculation of the third order
susceptibility still is a long
standing fundamental problem of particular importance in
nonlinear nanooptics: Indeed, 
cancellation of
size-dependent terms coming from
uncorrelated excitations is expected, but up to now shown for
very simple Hamiltonians only. Using a
many-body theory recently developed to handle interacting
close-to-bosons, we
prove it here for \emph{arbitrary} H. 
This new formalism actually
provides the first
clean way to calculate nonlinear
susceptibilities, with results
different from previous
ones.
\end{abstract}

\vspace{2cm}


\newpage

Size enhancement effect in nonlinear optical
processes is of great interest in nanoscience
and technology as it provides an
additional degree of freedom for the control
of matter. Care is however necessary in the
calculation of nonlinear susceptibilities due to possible
cancellation among their various terms [1].
Indeed, for non-interacting bosons, the cancellation is so
complete that all nonlinear susceptibilities vanish identically.
This shows that optical nonlinearities only come
from the non-bosonic nature of the excitations, and/or their
possible interactions.

In the various terms of the third order
susceptibility $\chi^{(3)}$, the ones with
uncorrelated excitations increase linearly
with the sample ``volume'' $L^D$. As the susceptibility
is an intensive quantity, all these
volume-linear terms have clearly to cancel. However, along
with them, other terms in $L^0$ can also disappear. This is why,
if this cancellation is not handled properly, the obtained
$\chi^{(3)}$ may well be incorrect. The cancellation problem is
thus crucial for nonlinear optics [2].

A lot of controversial arguments [3] have been raised to
justify the cancellation of volume-linear
terms. Some fundamental aspects were clarified through
the study of a simple model consisting of non-interacting 1D
Frenkel excitons with Pauli exclusion. It allows to analytically
show [4] the cancellation of these volume-linear terms for
arbitrary frequencies. Numerical evaluation of the
remaining terms shows a size
enhancement for sample sizes within the long wavelength
approximation and a possible saturation when the size approaches
the smaller of either, the coherence length due to non-radiative
scatterings, or the relevant wavelength of light [5]. This model
calculation, extended to higher dimension for excitons
having some kind of interactions [6], again shows cancellation of
the volume-linear terms.

Although these simple models give a reasonable picture of the
problem, they are not sufficient to carry out
reliable calculations in realistic situations. A general method
for calculating nonlinear susceptibilities which does not suffer
from this cancellation problem, was thus highly needed.

A new many-body theory has been recently
developed [7-13] to handle interactions between close-to-bosons:
It allows to extract from the quantities of physical
interest, the parts coming from pure Pauli ``interaction'' and
the parts coming from Coulomb interaction, either through direct
or exchange processes. We use this ``commutation
technique'' here, to solve the cancellation problem of the third
order susceptibility, by \emph{proving} that the terms
independent from exciton interactions cancel out exactly
\emph{for any matter Hamiltonian}. As expected, the remaining
terms of
$\chi^{(3)}$ depend on interactions only, pure Pauli or Coulomb
dressed by exchange, our new formalism allowing to write them in
a compact way.

The third order nonlinear susceptibility is the integral kernel
relating the field vector potentials to the current density
induced by the
semiconductor-photon coupling taken at third order. For photons
turned on adiabatically from
$-\infty$ to $t_0$, this induced current density reads [14]
\begin{equation}
\v
J =\int_{-\infty}^{t_0}dt_1\int_{-\infty}^{t_1}dt_2
\int_{-\infty}^{t_2}dt_3\,\langle
v|\,[\,[\,[\v I_0,I_1],I_2],I_3]|v\rangle\ ,
\end{equation}
where $I_{j\neq 0}=-ie^{iHt_j}H_{t_j}^{\mathrm{int}}e^{-iHt_j}$,
while $\v I_0=e^{iHt_0}\v J_{\v r}e^{-iHt_0}$. $H$ is
the matter Hamiltonian, $\v J_{\v r}$
the induced current density operator and
$H_t^\mathrm{int}= e^{\epsilon t}(e^{i\omega t}U+h.c.)$
the coupling between matter and $(\omega,\v Q)$ photons, if we
consider one photon field only for simplicity, $\epsilon=0_+$
being the adiabatic switching factor. $\v J_{\v r}$ and
$H_t^\mathrm{int}$ create or destroy one excitation in the
sample. In terms of creation
operators for the one-excitation eigenstates
$B_m^\dag$ defined as
$(H-E_m)B_m^\dag| v\rangle=0$, they formally read 
\begin{equation}
\v J_{\v r}=\sum_m\v J_{\v r,m}\ ,\hspace{0.5cm} \v J_{\v r,m}=\v
j _m(\v r)\,B_m+h.c.\ ,
\end{equation}
\begin{equation}
U\simeq\sum_m \mu_mB_m\ ,
\end{equation}
eq.\ (3) being valid within the rotating wave approximation. For
Wannier excitons, $m$ stands for $(\nu_m,\v Q_m)$ with
$\v Q_m$ being the exciton center of mass momentum and $\nu_m$
the relative motion index; $\mu_m=\v A.\v
G\,\delta_{\v Q_m,\v Q}\,\langle\v r=\v
0|\nu_m\rangle\,L^{D/2}$, while $\v j_m(\v r)=-\v G\,e^{i\v
Q_m.\v r}\,\langle\v r=\v 0|\nu_m\rangle\,L^{-D/2}$, with
 $\v A$ being the field vector potential and $\v G$ the
valence-conduction Kane vector [15].

Due to eq.\ (2), $\v J$ thus appears as 
$\sum_m
\v J_m$, where $\v J_m$ is given
by eq.\ (1), with $\v I_0$ calculated using 
$\v J_{\v r,m}$, instead of $\v J_{\v r}$.
By expanding the commutator of eq.\ (1) and by noting that
$I_{j\neq 0}^\dag=-I_j$ while $\v I_0^\dag=\v I_0$, 
$\v J_m$ actually reads $\tilde{\v J}_m+c.c.$, with
$\tilde{\v J}_m=
\v K_{0123}+\v K_{2103}+\v K_{3102}+\v K_{3201}$ and
\begin{equation}
\v K_{ijkl}=\int_{-\infty}^{t_0}dt_1\int_{-\infty}^{t_1}dt_2
\int_{-\infty}^{t_2}dt_3\,\langle v|I_i\,I_j\,I_k\,I_l|v\rangle
\ .
\end{equation}

\emph{The purpose of this letter} is to show how the 8 terms of
$\v J_m$ can be calculated formally, \emph{i}.\
\emph{e}., without knowing
$H$, in order \emph{(i) to prove that the sum of their
non-interacting contributions cancel out exactly, (ii) to
write a compact expression of the remaining terms}.

As explicitly shown below, the $\langle
v|I_i\,I_j\,I_k\,I_l|v\rangle$'s of eq.\ (4) split
into ``easy'' terms in which the middle state is
the vacuum --- so that they can be calculated exactly as they
depend on one-excitations only --- and ``tricky'' terms in which
the middle state contains two excitations. As the
two-excitation eigenstates are usually unknown, it has long
seemed hopeless to prove that, whatever $H$ is, these ``tricky''
terms contain parts which exactly cancel the ``easy'' terms. 

Using the ``commutation technique''[8,9] we recently developed to
handle many-body effects between close-to-bosons, we
can perform commutations between
$B_m^\dag$ and the $H$-dependent operators of these ``tricky''
terms, to rewrite them as a sum of
``non-interacting'', ``Pauli'' and
``Coulomb'' contributions. 
The ``tricky'' terms of the $\v K_{ijkl\neq 0123}$'s of
$\tilde{\v J}_m$ then read 
$\v j_m^\ast(\v
r,t_0)\,(\beta_{ijkl}+\gamma_{ijkl}+\delta_{ijkl})$, with
$\v j_m^\ast (\v r,t_0)=\v j_m^\ast(\v r)
e^{(3\epsilon+i\omega)t_0}$, the prefactors
$\beta_{ijkl}$, $\gamma_{ijkl}$, $\delta_{ijkl}$ coming from
non-interacting, Pauli and Coulomb processes, respectively.
$\v K_{0123}^{(\mathrm{tricky})}$ has a similar form, with
$\v j_m^\ast (\v r,t_0)$ replaced by $\v j_m(\v r,t_0)$, while
the ``easy'' terms of all the $\v K_{ijkl}$'s of
$\tilde{\v J}_m$ read
$\v j_m(\v r,t_0)\,\alpha_{ijkl}$. 

From the formal expressions of these
$\alpha$'s and $\beta$'s, which appear as
$H$ matrix elements, it is then possible to show the far from
obvious relations, $\alpha_{0123}+\alpha_{2103}+\beta_{3102}
^\ast=0$, $\alpha_{3201}+\alpha_{3102}+\beta_{2103}^\ast=0$ and
$\beta_{0123}+\beta_{3201}^\ast=0$, \emph{whatever $H$ is} [16].
Consequently, only remains in the induced current density, the
$\gamma$ and $\delta$ parts coming from Pauli and Coulomb
interactions between two excitations, as expected. It ultimately
reads 
\begin{eqnarray}
\v J_m=\v j_m^\ast(\v r)\, e^{i\omega
t_0}\,\frac{2}{E'_m}\,\sum_{p,q,n}\frac{\mu_p
\mu_q\mu_n^\ast}{E'_q\,E'_n}\hspace{3cm}\nonumber
\\ \times\left[\lambda_{pqmn}+\frac
{\xi_{pqmn}^\mathrm{dir}-\xi_{pqmn}^\mathrm{in}}{E'_p+E'_q}+\cdots
\right]\,+ c.c.\ ,
\end{eqnarray} 
$E'_i=E_i-\omega$ being the detuned exciton energy, the
next term having one more $\xi$ divided by detuning.
 
$\lambda_{pqmn}$, $\xi_{pqmn}^\mathrm{dir}$ and $\xi_{pqmn}
^\mathrm{in}$ are the exchange, direct Coulomb and ``in''
exchange Coulomb scatterings of the ``commutation technique''.
Defined through [8,9]
\begin{equation}
[B_m,B_i^\dag]=\delta_{mi}-D_{mi},\ \ \ [D_{mi},B_j^\dag]=
2\sum_n\lambda_{mnij}B_n^\dag,
\end{equation}
\begin{equation}
[H,B_i^\dag]=E_iB_i^\dag+V_i^\dag,\ \ \ [V_i^\dag,B_j^\dag]=
\sum_{mn}\xi_{mnij}^\mathrm{dir}B_m^\dag B_n^\dag,
\end{equation}
and $\xi_{mnij}^\mathrm{in}=\sum_{rs}\lambda_{mnrs}
\xi_{rsij}^\mathrm{dir}$, they read in terms of
the one-excitation eigenwavefunctions of $H$ (see eqs.\ (26,28)
of ref.\ [8]). The
``deviation-from-boson'' operator
$D_{mi}$ and the $\lambda_{mnij}$'s physically come from Pauli
exclusion which makes the excitons close-to-bosons only, while
the ``creation potential'' $V_i^\dag$ and the $\xi_{mnij}$'s come
from Coulomb interactions with the
$i$ exciton. 

To calculate
$\v J_m$, we have also used [10]
\begin{equation}
\frac{1}{a-H}\,B_m^\dag=\left(B_m^\dag+\frac{1}{a-H}\,V_m^\dag
\right)\,\frac{1}{a-H-E_m}\ ,
\end{equation}
which follows from eq.\ (7) and
\begin{equation}
e^{-iH\tau}B_m^\dag=B_m^\dag\,e^{-i(H+E_m)\tau}+W_m^\dag(\tau)\ ,
\end{equation}
\begin{equation}
W_m^\dag(\tau)=\int_{-\infty}^{+\infty}\frac{dx}{2i\pi}\,\frac
{e^{-i(x-i\eta)\tau}}{x-H-i\eta}\,V_m^\dag\,\frac{1}
{x-H-E_m-i\eta}\ ,
\end{equation}
which follows from eq.\ (8) and the integral representation of
$e^{-iH\tau}$ which, for $\tau<0$ and $\eta>0$, reads [17]
\begin{equation}
e^{-iH\tau}=\int_{-\infty}^{+\infty}\frac{dx}{2i\pi}\,\frac
{e^{-i(x-i\eta)\tau}}{x-H-i\eta}\ .
\end{equation}

To show how this ``commutation technique''
works and to grasp the physics which
controls the $\alpha$, $\beta$, $\gamma$, $\delta$ prefactors,
let us consider a $\v K_{ijkl}$ particularly complicated to
calculate, namely $\v K_{2103}$, because in addition to the fact
that its middle state has two excitations, the times $t_i$ are
``shaken up'' compared to $\v K_{0123}$, which makes the
integration over times more difficult.

$e^{-iHt}|v\rangle=|v\rangle$ and $U|v\rangle=0$, so that
$\langle v|I_2$ and $I_3|v\rangle$ reduce to one term. As
the non-zero contributions of $I_1\v I_0$ to $\v k_{2103}=\langle
v|I_2I_1\v I_0I_3|v\rangle$ are the ones in
$U^\dag B_m$ and $UB_m^\dag$ only, 
$\v k_{2103}$ splits in two. 
Its ``easy''
term, containing $U^\dag B_m$, reads $\v
k_{2103}^{(\mathrm{easy})}=i\,e^{\epsilon
(3\tau_1+2\tau_2+\tau_3)}\,a_{2103}\linebreak\v j_m(\v r,t_0)$
with
\begin{equation}
a_{2103}=\langle v|U\,e^{-iH'\tau_2}\,U^\dag\,e^{-iH
\tau_1}\,B_m\,e^{iH'(\tau_1+\tau_2+\tau_3)}\,U^\dag|v
\rangle\ ,
\end{equation}
where $H'=H-\omega$, while the $\tau_i$'s, defined as
$t_i=\tau_i+t_{i-1}$, run from $-\infty$ to $0$.
The $a_{2103}$ middle state being the vacuum,
the middle $H$ can be replaced by 0 and the right $H'$ by
$E'_m$. As $\langle v|B_mU^\dag|v\rangle=\mu_m^\ast$, the formal
integration of this ``easy'' term over the $\tau_i$'s readily
gives a
$\alpha_{2103}\,\v j_m(\v r,t_0)$ contribution to $\v K_{2103}$,
with
\begin{equation}
\alpha_{2103}=\frac{\mu_m^\ast}{(E'_m-3i\epsilon)(E'_m
-i\epsilon)}\ \langle v|U\,\frac{1}{H'-E'_m+2i\epsilon}\,
U^\dag|v\rangle\ .
\end{equation}

The $\v k_{2103}$ ``tricky'' term reads 
$\v k_{2103}^{(\mathrm{tricky})}=i\,e^{\epsilon
(3\tau_1+2\tau_2+\tau_3)}\,b_{2103}\,\v j_m^\ast(\v r,t_0)$ with
\begin{equation}
b_{2103}=\langle
v|U\,e^{-iH'\tau_2}\,U\,e^{-iH''\tau_1}\,B_m^\dag
\,e^{iH'(\tau_1+\tau_2+\tau_3)}\,U^\dag|v\rangle\ ,
\end{equation}
and $H''=H-2\omega$. The integrations over $\tau_1$ and $\tau_2$
are not straightforward because these times appear in different
places. We can put the
$\tau_1$'s together by passing $e^{-iH''\tau_1}$ over $B_m^\dag$
through eq.\ (9). This splits $b_{2103}$ as
$c_{2103}+d_{2103}$: From the first term
of eq.\ (9) we get
\begin{equation}
c_{2103}=e^{-iE'_m\tau_1}\,\langle v|U\,e^{-iH'\tau_2}
\, U\, B_m^\dag\,e^{iH'(\tau_2+\tau_3)}\,U^\dag|v\rangle\ .
\end{equation}
$\tau_2$ still is in two places. To go
further, we rewrite $UB_m^\dag$ as $B_m^\dag U+[U,B_m^\dag]$,
\emph{i}.\ \emph{e}., $B_m^\dag U+\mu_m-\sum_q\mu_qD_{qm}$,
due to eqs.\ (3,6). In the part with $B_m^\dag U$, we can replace
the left $H'$ by $E'_m$ since the middle state is then
the vacuum. As $\langle v|UB_m^\dag|v\rangle=\mu_m$,
the two first terms of $UB_m^\dag$ give a $\beta_{2103}\,\v
j_m^\ast(\v r,t_0)$ contribution to
$\v K_{2103}$, with
\begin{eqnarray}
\beta_{2103}=\frac{\mu_m}{E'_m+3i\epsilon}\hspace{6cm}
\nonumber
\\ \times \langle
v|\,U
\left(\frac{1}{H'-E'_m-2i\epsilon}-\frac{1}{2i\epsilon}
\right)\frac{1}{H'-i\epsilon}\,U^\dag\,|v\rangle\ .
\end{eqnarray}

As for the third term of $UB_m^\dag$, in $D_{qm}$, it
physically comes from Pauli ``interaction'', so that it is
zero for boson-excitons. The contribution it induces to
$\v K_{2103}$ reads $\gamma_{2103}\,\v j_m^\ast(\v
r,t_0)$ with
\begin{eqnarray}
\gamma_{2103}=\sum_q\frac{-i\mu_q}{E'_m+3i\epsilon}\ \int
_{-\infty}^{0}d\tau_2\,e^{2\epsilon\tau_2}\hspace{1cm}
\nonumber
\\ \times \langle
v|\,U\,e^{-i H'\tau_2}\,D_{qm}\,\frac{e^{iH'\tau_2}}{H'-i
\epsilon}\, U^\dag\,|v\rangle\ .
\end{eqnarray}

Since $H'$ acts on one-excitations only, the
integration over $\tau_2$ can be performed exactly, using eq.\
(3). As $\langle
v|B_pD_{qm}B_n^\dag|v\rangle=2\lambda_ {pqmn}$,
due to eq.\ (6), we find [18]
\begin{equation}
\gamma_{2103}=-\,\frac{2}{E'_m+3i\epsilon}\,\sum_{p,q,n}\frac
{\lambda_{pqmn}\,\mu_p\,\mu_q\,\mu_n^\ast}{(E'_n-i\epsilon)
(E'_n-E'_q-2i\epsilon)}\ .
\end{equation}

In passing $e^{-iH''\tau_1}$ over $B_m^\dag$ in eq.\
(14), we also generate a $d_{2103}$ term
which comes from $W_m^\dag(\tau_1)$ in eq.\ (9), \emph{i}.\
\emph{e}., Coulomb interactions with the $m$ exciton. This term
produces a
$\delta_{2103}\,\v j_m^\ast(\v r,t_0)$ contribution to $\v
K_{2103}$ with
\begin{equation}
\delta_{2103}=i\int_{-\infty}^{0}d\tau_2\,e^{2\epsilon\tau_2}\,
\langle v|\,U\,e^{-iH'\tau_2}\,U\,\overline{W}_m^\dag\,
\frac{e^{iH'\tau_2}}{H'-i\epsilon}\,U^\dag\,|v\rangle\ ,
\end{equation}
where $\overline{W}_m^\dag$ is linked to $W_m^\dag(\tau_1)$
through
\begin{equation}
\overline{W}_m^\dag=-i\int_{-\infty}^{0}d\tau_1\,W_m^\dag(\tau_1)
\,e^{(3\epsilon+i(H'+2\omega))\tau_1}\ .
\end{equation}
Using eq.\ (10), it reads
\begin{eqnarray}
\overline{W}_m^\dag=
\int_{-\infty}^{+\infty}\frac{dy}{2i\pi}\,\frac{1}
{y-H''-i\eta}\,V_m^\dag\hspace{2cm}\nonumber
\\ \times\frac{1}{(y-H'+i\epsilon
')(y-H'-E'_m-i\eta)}\ ,
\end{eqnarray}
with $\eta$ chosen such that $\epsilon '=3\epsilon-\eta>0$, to
insure convergence for
$\tau_1\rightarrow -\infty$.

The integration over $\tau_2$  can be
performed exactly, using again eq.\ (3). This leads to [18]
\begin{equation}
\delta_{2103}=\frac{1}{E'_m+3i\epsilon}\,\sum_{p,q,n}
\frac{w_{pqmn}\
\mu_p\,\mu_q\,\mu_n^\ast}{(E'_n-i\epsilon)(E'_n-E'_q-2i\epsilon)}
\ .
\end{equation}
$w_{pqmn}=(E'_m+3i\epsilon)\,\langle
v|B_pB_q\overline{W}_m^\dag B_n^\dag|v\rangle$ corresponds to
Coulomb processes between the $m$ exciton and the photocreated
ones $(p,q)$ and $n$. When inserted in $w_{pqmn}$, the
$\overline{W}_m^\dag$, of eq.\ (21) can have its
two right $H'$ replaced by $E'_n$. Integration over $y$, done by
residues, yields
\begin{equation}
w_{pqmn}=\langle v|B_p\,B_q\,\frac{1}{
E'_n-H''-3i\epsilon}\,V_m^\dag\,B_n^\dag|v\rangle\ .
\end{equation}
$w_{pqmn}$ cannot be calculated exactly because
$H''$ acts on two excitons. It can however be expanded 
in Coulomb interactions using eq.\ (8). At lowest order, this
leads to replace $H''$ by
$E'_p+E'_q$. From eq.\ (7) and the scalar
product of two-exciton states,
$\langle v|B_pB_qB_r^\dag
B_s^\dag|v\rangle=\delta_{pr}\delta_{qs}+\delta_{ps}\delta_{qr}
-2\lambda_{pqrs}$, which follows from eq.\ (6), we end with
\begin{equation}
w_{pqmn}\simeq
2\,\frac{\xi_{pqmn}^\mathrm{dir}-\xi_{pqmn}^\mathrm{in}}
{E'_n-E'_p-E'_q-3i\epsilon}+\cdots\ .
\end{equation}

$\xi_{pqmn}^\mathrm{in}$ is an exchange
Coulomb scattering
with its Coulomb
interactions between the ``in'' excitons
$(m,n)$ only. Once again [8,9,10], it differs from the ``out''
exchange Coulomb scattering
$\xi_{pqmn}^\mathrm{out}=
\sum_{r,s}\xi_{pqrs}^\mathrm{dir}\,\lambda_{rsmn}$ appearing in
the effective bosonic Hamiltonian for excitons [19], widely used
up to now, in spite of the fact that it is unphysical because
not hermitian.

The ``commutation technique'' being new, and
the project to extract from $\v J$ its
non-interacting terms ambitious, it seemed to us useful
to detail the calculation of a particular $\v K_{ijkl}$, even if
this calculation may appear somewhat technical. The 
expressions of $\alpha$,
$\beta$,
$\gamma$,
$\delta$'s it provides, are actually quite interesting.

(i) The $\alpha$'s and $\beta$'s, independent from 
Pauli and Coulomb interactions, contain three exciton-photon
couplings
$\mu_p\mu_q\mu_n^\ast$. In the case of Wannier excitons, they
generate three factors
$L^{D/2}$ and force $\v Q_p=\v Q_q=\v Q_n=\v Q$ so that no
sample volume comes from the sums over $(p,q,n)$.
By including
the $L^{-D/2}$ factor of $\v j_m(\v r)$, these $\alpha$'s and
$\beta$'s thus generate  volume-linear contributions
$L^D$ to $\v J_m$, in agreement with model Hamiltonians. They
have to and do
cancel exactly for any kind of excitons.

(ii) The $\gamma$'s and $\delta$'s have the
same number of exciton-photon couplings. They however have one
additional Pauli ``scattering''
for $\gamma$'s, and Coulomb
scattering for
$\delta$'s. As, for Wannier excitons [8,9,12], these
scatterings both behave as
the exciton volume divided by $L^D$, the
$\gamma$ and $\delta$ contributions to the current density are
indeed sample volume free. 

(iii) Finally, at large
detuning, the induced current density is dominated by pure Pauli
processes: As obvious from eqs.\ (6,7), the
$\lambda_{pqmn}$'s are dimensionless, while the $\xi_{pqmn}$'s
have the dimension of an energy, so that the $\delta$'s do have
one more energy denominator than the $\gamma$'s, which makes them
smaller in the large detuning limit.

\noindent\emph{Conclusion}

Using a new many-body theory
for interacting close-to-bosons, we have succeeded in
extracting the various parts of the induced current density
linked to uncorrelated excitations and proved their cancellation,
\emph{without knowing $H$}. This formalism actually provides
\emph{the first clean way to calculate nonlinear
susceptibilities.} After combining all its contributions,
the induced current density appears in terms
of the exchange and Coulomb ``scatterings'' of the ``commutation
technique'' (see eq.\ (5)). Explicit calculations for Wannier and
Frenkel excitons, as well as polarization effects linked to spin
degrees of freedom, will be presented elsewhere. We can however
say, just from dimensional arguments, that, at large detuning, 
nonlinear susceptiblities are entirely controlled by Pauli
interactions between excitations, without the help of any Coulomb
process --- result beyond the reach of the effective bosonic
Hamiltonians for excitons.

\vspace{0.3cm}

This work was supported by the CNRS of France and by the
Grant-in-Aid for Scientific Research (No. 15540311) of the
Ministery of Education, Culture, Sports, Science and Technology
of Japan.
\vspace{0.3cm}

\hbox to \hsize {\hfill REFERENCES
\hfill}

\noindent
(1) L. Banyai, Y.Z. Hu, M. Lindberg, S.W. Koch, Phys.\ Rev.\ B
\underline{38}, 8142 (1988)

\noindent
(2) K. Cho, ``Optical Response of Nanostructures: Microscopic
Nonlocal Theory'', (Springer Verlag, 2003), Sec.3.9

\noindent
(3) E. Hanamura, Solid State Com.\ \underline{62}, 465 (1987);
\emph{ibid}.\ Phys.\ Rev.\ B \underline{37}, 1273 (1988); L.
Banyai et al., ref.\ (1); F.C. Spano, S. Mukamel, Phys.\ Rev.\ A
\underline{40}, 5783 (1989)

\noindent
(4) H. Ishihara, K. Cho, Phys.\ Rev.\ B \underline{42}, 1724
(1990)

\noindent
(5) H. Ishihara, K. Cho, Int.\ J.\ Nonlin.\ Opt.\ Phys.\
\underline{1}, 287, (1992)

\noindent
(6) H. Ishihara, T. Amakata, Int.\ J.\ Mod.\ Phys.\ B
\underline{15}, 3809 (2001); H. Ishihara, T. Nakatani, CLEO/QELS
Tech.\ Digest, CD-ROM (2003, Baltimore) QTuG13

\noindent
(7) M. Combescot, C. Tanguy, Europhys.\ Lett.\
\underline{55}, 390 (2001)

\noindent
(8) M. Combescot (M.C.), O. Betbeder-Matibet (O.B.M.),
Europhys.\ Lett.\ \underline{58}, 87 (2002)

\noindent
(9) O.B.M., M.C., Eur.\ Phys.\ J.\ B \underline{27}, 505 (2002)

\noindent
(10) M.C., O.B.M., Europhys.\ Lett.\ \underline{59}, 579 (2002)

\noindent
(11) M. C., X. Leyronas, C. Tanguy, Eur.\ Phys.\ J.\ B
\underline{31}, 17 (2003)

\noindent
(12) O.B.M., M.C., Eur.\ Phys.\ J.\ B, \underline{31}, 517 (2003)

\noindent
(13) M.C., O.B.M., Solid State Com.\ \underline{126}, 687 (2003)

\noindent
(14) See e.g., eq.\ (4) of ref.\ (1) or eq.\ (2.109) of ref.\ (2)

\noindent
(15) See for instance, O.B.M., M.C., Eur.\ Phys.\ J.\ B
\underline {22}, 17 (2001)

\noindent
(16) This grouping was already noted in ref.\ (4) and ascribed
to the classification of the various terms, according to their
frequency dependences, which guarantees a cancellation
independent from frequencies.

\noindent
(17) See for instance, M.C., J.\ Phys.\ A \underline{34}, 6087
(2001)

\noindent
(18) $\gamma_{2103}$ and $\delta_{2103}$ seem to have singular
terms for $q=n$. They do in fact combine with similar ones
appearing in $\gamma_{3102}$ and $\delta_{3102}$ to ultimately
give eq.\ (5). 

\noindent
(19) See for example H. Haug, S. Schmitt-Rink, Progr.\ Quantum.\
Electron.\ \underline{9}, 3 (1984)

\end{document}